\documentclass[12pt,lettersize,journal,onecolumn]{IEEEtran}
\usepackage{amsmath,amsfonts}
\usepackage{algorithmic}
\usepackage{algorithm}
\usepackage{array}
\usepackage[caption=false,font=normalsize,labelfont=sf,textfont=sf]{subfig}
\usepackage{textcomp}
\usepackage{stfloats}
\usepackage{url}
\usepackage{verbatim}
\usepackage{graphicx}
\def\BibTeX{{\rm B\kern-.05em{\sc i\kern-.025em b}\kern-.08em
    T\kern-.1667em\lower.7ex\hbox{E}\kern-.125emX}}
\usepackage{balance}
\usepackage{setspace}
\usepackage[
colorlinks=true,
citecolor=blue,
linkcolor=blue,  
urlcolor=blue
]{hyperref}
\begin{document}
\title{Sensing-Driven Secure Transmission with Mobile Eavesdropper Tracking in RIS-Assisted ISAC}
\author{
	Jie Li,
	Ning Wang,
	Zhendong Li,
	Ze Li,
	Lingfeng Shen,
	Zhongyun Li,
	and Jiakang Zheng
	\thanks{Jie Li, Ning Wang and Ze Li are with the School of Electrical and Information Engineering, Zhengzhou University, Zhengzhou 450001, China (e-mail: 15534188602@163.com; lize0802@gs.zzu.edu.cn; ienwang@zzu.edu.cn). Zhendong Li is with the School of Information and Communication Engineering, Xi’an Jiaotong University, Xi’an 710049, China (e-mail: lizhendong@xjtu.edu.cn). 
	Lingfeng Shen is with the School of Software, Henan University, Zhengzhou 450046, China (e-mail: ielfshen@henu.edu.cn).
	Zhongyun Li is with the School of Electronic and Information Engineering, Tongji University, Shanghai 201804, China (e-mail: zhongyun\_li@tongji.edu.cn).
	Jiakang Zheng is with the School of Electronic and Information Engineering, Beijing Jiaotong University, Beijing 100044, China (e-mail: jiakangzheng@bjtu.edu.cn).
		
	(Corresponding author: Zhendong Li)}
	}

\maketitle

\begin{abstract}
	The mobility of potential eavesdroppers (Eves) poses a critical challenge to physical-layer security (PLS) in integrated sensing and communication (ISAC) systems, since rapidly varying Eve positions lead to time-varying eavesdropping channels and outdated security decisions. To address this challenge, this paper investigates sensing-driven secure transmission for reconfigurable intelligent surface (RIS)-assisted ISAC systems with a maneuvering Eve. To this end, by exploiting the dual functionality of the ISAC waveform, a multi-view angle estimation method jointly exploiting the base station (BS) and RIS is developed based on MUSIC and maximum likelihood estimation (MLE), enabling three-dimensional (3D) Eve localization, whose accuracy is characterized by the Cramér-Rao lower bound (CRLB). Subsequently, an interacting multiple model extended Kalman filter (IMM-EKF) with out-of-sequence measurement (OOSM) compensation is employed to track and predict the maneuvering Eve. Based on the predicted trajectory, the time-varying eavesdropping channel is updated, and the transmit beamforming and RIS phase shifts are jointly optimized subject to a sensing CRLB constraint. Furthermore, an alternating optimization (AO) algorithm integrating fractional programming, semidefinite relaxation (SDR), and Riemannian manifold optimization is developed to solve the resulting non-convex problem. Finally, numerical results demonstrate the effectiveness of the proposed approach in accurately tracking maneuvering Eves and enhancing secure transmission in dynamic ISAC environments.
\end{abstract}

\begin{IEEEkeywords}
Integrated sensing and communication (ISAC), reconfigurable intelligent surface (RIS), physical-layer security, mobile eavesdropping, trajectory tracking.
\end{IEEEkeywords}

\section{Introduction}
\IEEEPARstart{W}ith the rapid development of sixth-generation (6G) wireless networks, future wireless systems are expected to provide high-rate and reliable communications while supporting intelligent sensing of the surrounding environment~\cite{8869705,9040264}. Integrated sensing and communication (ISAC) emerged as a key technology to meet these requirements by enabling communication and sensing within a unified wireless system. In particular, ISAC allows communication infrastructures to simultaneously deliver information and acquire environmental information through the sharing of spectrum, hardware, and signal processing resources~\cite{9737357,9540344}. This integrated design can improve spectrum and hardware utilization while reducing the need for dedicated sensing infrastructures. As a result, ISAC attracted increasing attention in a variety of applications, including autonomous systems, intelligent transportation, and other dynamic wireless environments, where timely environmental awareness is essential for reliable communication and intelligent decision making~\cite{8999605}.

Despite these advantages, the broadcast nature of wireless transmission makes ISAC systems vulnerable to eavesdropping attacks~\cite{9705498}. In conventional physical-layer security (PLS) schemes, the secrecy performance depends heavily on the availability of accurate channel state information (CSI) of potential eavesdroppers (Eves)~\cite{5605343}. In practice, however, obtaining accurate Eve CSI is considerably more difficult than acquiring the legitimate user's CSI, since the Eve is typically a passive receiver and does not actively participate in channel estimation. This requirement becomes particularly challenging in dynamic scenarios, where an Eve may be mounted on a mobile platform, such as an unmanned aerial vehicle (UAV) or a mobile vehicle, causing its channel to vary continuously with its motion~\cite{8618602}. The resulting channel aging effect can make previously acquired Eve information increasingly mismatched with the actual eavesdropping channel, especially when the Eve moves at a relatively high speed or performs maneuvering motions. As a result, secure transmission strategies developed based on static or outdated Eve information can suffer from substantial performance degradation~\cite{8618602,6482662}. Therefore, acquiring timely and accurate information about the Eve's position and motion state, and further incorporating such information into adaptive secure transmission design, is essential for maintaining reliable security in dynamic wireless environments.

To acquire timely information about a mobile Eve and adapt the wireless transmission accordingly, it is important to introduce additional controllable propagation mechanisms into ISAC systems. Recently, reconfigurable intelligent surface (RIS) attracted considerable attention for its ability to reconfigure wireless propagation environments~\cite{8910627,8796365}. By adjusting the phase shifts of its reflecting elements, an RIS can reshape propagation paths, enhance desired communication signals, and mitigate potential information leakage~\cite{8910627,9014322,11194496}. More importantly, the controllable reflection of the RIS can create additional propagation paths for ISAC sensing, providing complementary spatial observations of a target through BS-RIS-Eve reflection links~\cite{9508872}. Such multi-view sensing capability is particularly useful for acquiring the position and motion information of a mobile Eve, while the reconfigurable reflection can be further optimized to improve secure transmission. In addition to reflective RIS architectures, transmissive RIS-based transceivers were also recently investigated for low-cost and low-power communication systems, covering both system architecture design and efficient beamforming optimization~\cite{10680462,10711860}. Moreover, the spatial degrees of freedom introduced by RIS can facilitate the joint adaptation of sensing and communication functions, allowing the propagation environment to be dynamically configured according to the acquired target information. This capability is particularly attractive in dynamic scenarios, where the sensing configuration and secure transmission strategy need to be continuously updated as the Eve moves. Therefore, RIS-assisted ISAC provides a natural platform for integrating Eve sensing with adaptive physical-layer security in dynamic wireless environments.

Despite the advantages of RIS assisted ISAC, existing studies on physical layer security mainly focused on static or quasi static scenarios. Several works investigated secure transmission in RIS assisted ISAC systems by jointly optimizing the transmit beamforming and RIS phase shifts to maximize the secrecy rate or other security related metrics~\cite{9594786,9420018}.  
Recent studies further investigated secure ISAC designs with more sophisticated transmission strategies. In particular, artificial noise was introduced to degrade the received signal quality at potential Eves, leading to the joint optimization of communication beamforming, artificial-noise transmission, and RIS phase shifts~\cite{10547419}. Other works considered multi-user or multi-target scenarios, where the active beamforming at the BS and passive beamforming at the RIS are jointly designed to balance communication quality, sensing performance, and information leakage~\cite{Elsayed2025SumSecrecy}. These studies demonstrated that the additional spatial degrees of freedom provided by the RIS can be effectively exploited to enhance physical-layer security in ISAC systems.  
Other studies incorporated sensing requirements into the optimization process and explored the tradeoff between sensing accuracy and secure communication performance~\cite{Jiang2025RobustSecure}. These works demonstrate the potential of RIS assisted ISAC for improving both sensing and physical layer security. However, most of them assume that the locations or channel state information of potential Eves are perfectly known or remain unchanged during the transmission process.

To address the dynamic nature of potential Eves, some recent studies considered target localization and tracking based on sensing observations~\cite{1039398,8741198}. Recent studies further explored sequential target tracking and multiple-model filtering to handle time-varying and maneuvering target motion~\cite{Mazor1998IMM,BarShalom1989IMM}. In these approaches, estimation errors caused by noise, limited snapshots, and imperfect parameter extraction can affect the accuracy of the acquired target information. Moreover, sensing and tracking inevitably involve processing and measurement delays, such that the latest available observation may not accurately reflect the current Eve position. OOSM processing techniques were therefore developed to incorporate delayed observations into target-state estimation, including OOSM-aware IMM tracking for maneuvering targets~\cite{BarShalom2004OOSM,BarShalom2005IMMOOSM}. As a result, directly using the estimated Eve state for secure transmission can lead to outdated channel information, particularly when the Eve exhibits rapid or maneuvering motion. Recent sensing-assisted predictive beamforming studies showed that predicted target states can be exploited to proactively adapt transmission beams in time-varying wireless environments~\cite{Yuan2020BayesianPredictive,Xia2024SensingEnabledPB}. Although existing studies separately investigated RIS assisted secure transmission and target tracking, how to integrate sensing uncertainty, delayed observations, and Eve motion prediction into the joint design of beamforming and RIS phase shifts remains largely unexplored. This motivates the development of a sensing driven secure transmission scheme that can continuously exploit sensing and tracking information to adapt to the time varying Eve channel.

To address these challenges, this paper investigates sensing driven secure transmission for RIS assisted ISAC systems with a mobile Eve. The downlink communication waveform is reused for sensing, while different RIS reflection configurations provide additional spatial observations for Eve localization. The obtained sensing information is then used to estimate and predict the Eve's motion, thereby updating the time varying eavesdropping channel for secure transmission design. Based on the predicted Eve state, the BS beamforming vector and RIS phase shifts are jointly optimized under communication and sensing requirements. The resulting non convex optimization problem is solved using an alternating optimization (AO) approach. The main contributions of this work are summarized as follows.

\begin{itemize}
	\item Dynamic sensing and secure transmission modeling: A RIS assisted ISAC system with a mobile Eve is considered, where the downlink communication waveform is reused for sensing and multiple RIS reflection configurations are employed to obtain diverse observations of the Eve. A three dimensional (3D) localization model is established by jointly exploiting the angular information from the direct and RIS assisted paths, and the localization accuracy is characterized through a Cramér-Rao lower bound (CRLB) analysis. Considering sensing latency and Eve mobility, the acquired sensing information is further used to track and predict the Eve state, based on which a time varying eavesdropping channel is constructed. A secrecy rate maximization problem is then formulated by jointly optimizing the BS transmit beamforming and RIS phase shifts under communication quality of service and sensing accuracy constraints.
	
	\item Efficient optimization and tracking algorithms: To obtain accurate Eve state information, MUSIC and maximum likelihood estimation (MLE) based estimators are developed for extracting the angular parameters from different sensing paths, while an interacting multiple model extended Kalman filter (IMM-EKF) with out of sequence measurement (OOSM) compensation is employed to handle sensing delays and different Eve motion patterns. For the formulated secure transmission problem, an AO algorithm is developed to decouple the coupled design variables. Semidefinite relaxation (SDR) and manifold based optimization are further employed to efficiently solve the resulting beamforming and RIS phase shift subproblems, and the convergence and computational complexity of the proposed algorithm are analyzed.
	
	\item Comprehensive performance evaluation: Extensive numerical simulations are conducted to evaluate the proposed sensing and secure transmission design from multiple aspects. The localization performance is first verified through comparisons with benchmark angle estimation and localization methods, and the derived CRLB is used to characterize the impact of key system parameters. The tracking performance of the IMM-EKF with OOSM compensation is then evaluated under different Eve motion patterns and sensing delays. Finally, the secrecy performance of the proposed adaptive transmission strategy is compared with representative benchmarks, demonstrating the benefits of exploiting sensing and trajectory prediction for secure transmission against mobile Eves.
\end{itemize}

The remainder of this paper is organized as follows. The system model and channel assumptions are introduced in Section II. Section III presents the RIS-assisted sensing, localization, CRLB analysis, and delay-aware tracking framework. The sensing-driven secure transmission optimization algorithm is developed in Section IV. Simulation results are provided in Section V to verify the effectiveness of the proposed framework. Finally, conclusions are drawn in Section VI.

\subsubsection*{Notation}
Throughout this paper, bold lowercase and bold uppercase letters denote vectors and matrices, respectively. $(\cdot)^T$, $(\cdot)^H$ and $(\cdot)^*$ denote the transpose, conjugate transpose and conjugate, respectively. $\operatorname{tr}(\mathbf{A})$ and $\|\mathbf{A}\|_F$ denote the trace and Frobenius norm of matrix $\mathbf{A}$, respectively. $\operatorname{diag}(\mathbf{a})$ denotes a diagonal matrix whose diagonal entries are the elements of vector $\mathbf{a}$. $\mathbf{I}$ denotes the identity matrix of appropriate size. $\mathbb{C}$, $\mathbb{R}$ and $\mathbb{Z}$ denote the sets of complex, real and integer numbers, respectively. $\mathcal{CN}(\boldsymbol{\mu},\mathbf{R})$ denotes the circularly symmetric complex Gaussian distribution with mean $\boldsymbol{\mu}$ and covariance matrix $\mathbf{R}$. $\operatorname{Re}(\cdot)$ and $\operatorname{Im}(\cdot)$ denote the real and imaginary parts, respectively. $\mathbb{E}[\cdot]$ denotes statistical expectation. For a complex scalar, vector or matrix, $\|\cdot\|$ without subscript denotes the Euclidean norm.
\section{System Model}

\begin{figure}[!t]
	\centering
	\includegraphics[width=4.5in]{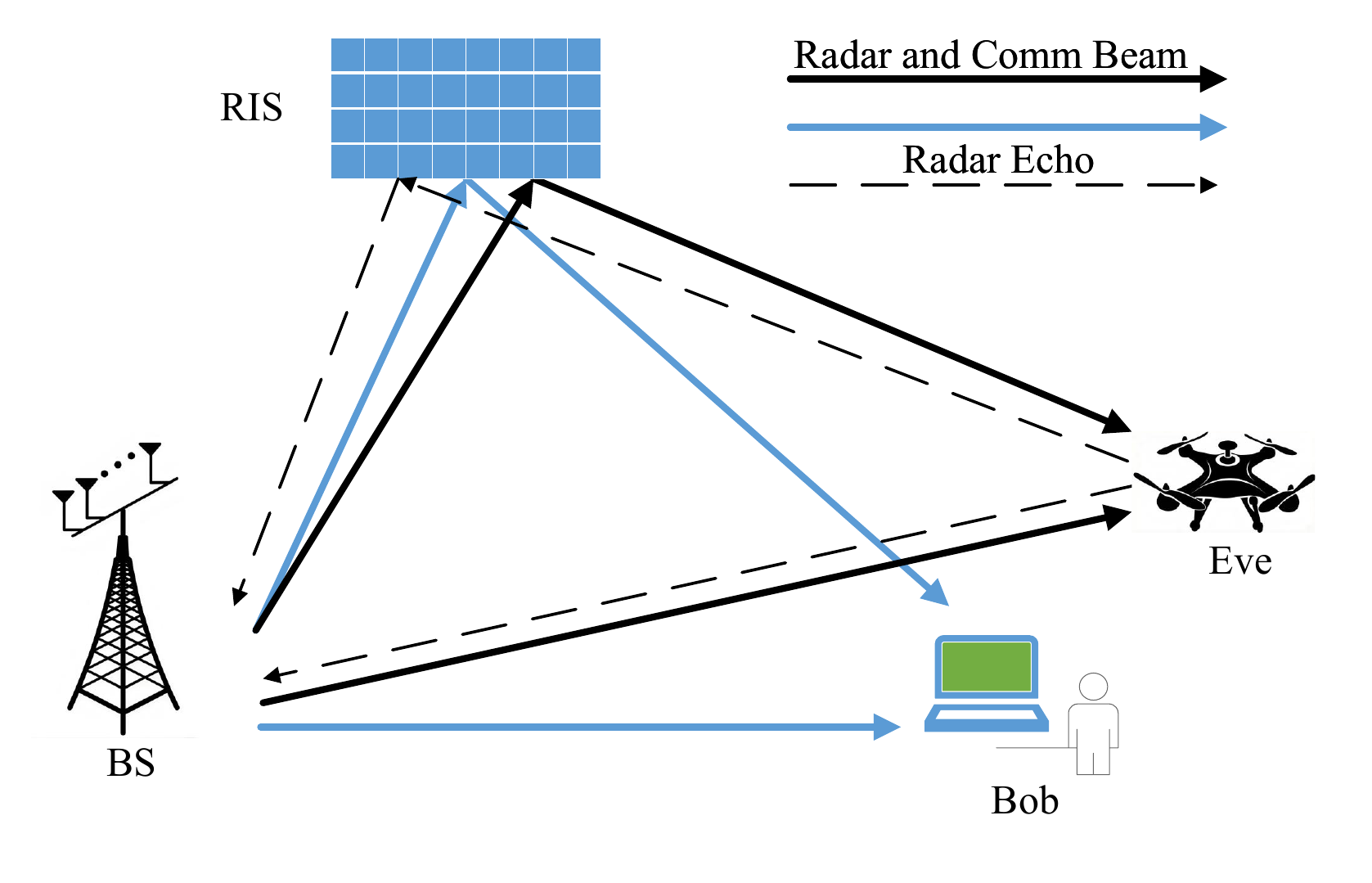}
	\caption{RIS-enabled secure ISAC system with a mobile Eve.}
	\label{Fig_1}
\end{figure}

As illustrated in Fig. \ref{Fig_1}, a RIS-assisted 3D secure ISAC system is considered. The system consists of a multi-antenna BS equipped with $M$ antennas, a RIS consisting of $N$ reflecting elements, and a single-antenna legitimate user (Bob). In addition, a malicious UAV, acting as a potential Eve, moves in the 3D space to intercept confidential information transmitted from the BS to Bob.
Due to the high mobility of the UAV, its instantaneous location is unknown and the corresponding channel varies over time. Therefore, the BS exploits the sensing information and motion prediction results to obtain the estimated Eve channel and further optimize the secure transmission strategy.

A composite channel model incorporating large-scale path loss and small-scale fading is adopted~\cite{8741198}. The large-scale fading coefficient of the $i$-th link is defined as
\begin{equation}
	\beta_i=C_0 d_i^{-\alpha_i},
\end{equation}
where $C_0$ denotes the reference path loss coefficient, $d_i$ represents the corresponding link distance, and $\alpha_i$ is the path loss exponent.
The small-scale fading follows the Rician distribution, consisting of line-of-sight (LoS) and non-line-of-sight (NLoS) components~\cite{8796365,8741198}. The NLoS components are modeled as independent and identically distributed (i.i.d.) circularly symmetric complex Gaussian random variables with zero mean and unit variance. Specifically, the random vectors are denoted by $\mathbf g_{dB},\mathbf g_{dE}\in\mathbb C^{M\times1}$ and $\mathbf g_{rB},\mathbf g_{rE}\in\mathbb C^{N\times1}$, and the NLoS component of the BS-RIS channel is denoted by $\mathbf G_{BR}^{\rm NLoS}\in\mathbb C^{N\times M}$, where
$\mathbf g_{dB}, \mathbf g_{dE} \sim \mathcal{CN}(\mathbf 0, \mathbf I_M)$,
$\mathbf g_{rB}, \mathbf g_{rE} \sim \mathcal{CN}(\mathbf 0, \mathbf I_N)$,
and each element of $\mathbf G_{BR}^{\rm NLoS}$ is independently distributed as $\mathcal{CN}(0,1)$.
Both the BS and RIS employ uniform planar arrays (UPAs). The
corresponding array response vectors are denoted by
$\mathbf a_{\rm B}(\theta,\phi)$ and
$\mathbf a_{\rm R}(\theta,\phi)$,
respectively. Here, $\theta$ and $\phi$ denote the azimuth and elevation
angles of the impinging wave, respectively.
The channel vectors of the BS-to-user links are defined as
\begin{equation}
	\mathbf h_i
	=
	\sqrt{\beta_i}
	\left(
	\sqrt{\frac{\mathcal K_i}{\mathcal K_i+1}}
	\mathbf a_{\rm B}(\theta_i,\phi_i)
	+
	\sqrt{\frac{1}{\mathcal K_i+1}}
	\mathbf g_i
	\right),
\end{equation}
where $i\in\{dB,dE\}$ denotes the direct BS-to-Bob and
BS-to-Eve links, respectively. Here, $\mathcal K_i$ denotes the Rician
factor.
The channel vectors of the RIS-to-user links are defined as
\begin{equation}
	\mathbf h_i
	=
	\sqrt{\beta_i}
	\left(
	\sqrt{\frac{\mathcal K_i}{\mathcal K_i+1}}
	\mathbf a_{\rm R}(\theta_i,\phi_i)
	+
	\sqrt{\frac{1}{\mathcal K_i+1}}
	\mathbf g_i
	\right),
\end{equation}
where $i\in\{rB,rE\}$ denotes the RIS-to-Bob and RIS-to-Eve links,
respectively.
The BS-RIS channel is modeled as
\begin{equation}
	\mathbf G_{BR}
	=
	\sqrt{\beta_{BR}}
	\left(
	\sqrt{\frac{\mathcal K_{BR}}{\mathcal K_{BR}+1}}
	\mathbf G_{BR}^{\rm LoS}
	+
	\sqrt{\frac{1}{\mathcal K_{BR}+1}}
	\mathbf G_{BR}^{\rm NLoS}
	\right),
\end{equation}
where
$\mathbf G_{BR}^{\rm LoS}=\mathbf a_{\rm R}(\theta_{RB},\phi_{RB})\mathbf a_{\rm B}^{H}(\theta_{BR},\phi_{BR})$,
$(\theta_{BR},\phi_{BR})$ and
$(\theta_{RB},\phi_{RB})$ denote the departure angle at the BS and the
arrival angle at the RIS, respectively.
Since the BS and RIS are deployed at fixed locations, the BS-RIS channel remains quasi-static and can be accurately estimated~\cite{9771077}. Meanwhile, Bob is assumed to be cooperative, and its CSI can be acquired through conventional pilot-based channel estimation~\cite{9014322}.
The RIS phase-shift matrix is defined as
$\mathbf{\Phi}=\operatorname{diag}(\mathbf v)=\operatorname{diag}(v_1,\ldots,v_N)$,
where
$\mathbf v=[v_1,\ldots,v_N]^T,v_n=e^{j\varphi_n},|v_n|=1,\ n=1,\ldots,N $.
Here, $\varphi_n\in[0,2\pi)$ denotes the adjustable phase shift introduced by the $n$-th RIS reflecting element~\cite{8910627}. The unit-modulus constraint indicates that the RIS only changes the phase of the incident signal while maintaining constant reflection amplitude.
The instantaneous Eve channel is unavailable due to its non-cooperative behavior and mobility. Therefore, $\mathbf h_E$ is obtained from the sensing-assisted location estimation and subsequent channel reconstruction.

\section{Radar Sensing and Angular Parameter Estimation}
To estimate the position of the mobile Eve, the BS exploits the downlink
communication waveform for radar sensing. During sensing intervals, the RIS
adopts different phase configurations to generate spatial diversity for
angular estimation~\cite{9456027}.
The considered system operates in a time division duplex (TDD) mode. Therefore, the uplink echo
channel from Eve to the BS is assumed to satisfy channel reciprocity with
the downlink BS-Eve channel. The transmitted signal is expressed as follows
\begin{equation}
	\mathbf{s}_{\mathrm{tx}}(t)
	=
	\mathbf{w}s(t),
\end{equation}
where $\mathbf{w}\in\mathbb{C}^{M\times1}$ is the transmit beamforming vector
satisfying $\|\mathbf{w}\|^2 \leq P_t$, and $s(t)$ is a unit-energy waveform.
For the $q$-th sensing interval, the equivalent BS-Eve channel is given as follows
\begin{equation}
	\mathbf h_{\mathrm{BE}}^{(q)}
	=
	\mathbf h_{dE}
	+
	\mathbf G_{BR}^{H}
	(\mathbf\Phi^{(q)})^{H}
	\mathbf h_{rE},
\end{equation}
where $\mathbf h_{dE}$ denotes the direct BS-Eve channel,
$\mathbf G_{BR}$ represents the BS-RIS channel,
$\mathbf h_{rE}$ denotes the RIS-Eve channel, and
$\mathbf\Phi^{(q)}$ denotes the RIS phase-shift matrix.
Therefore, the received echo signal at the BS is expressed as
\begin{equation}
	\mathbf y^{(q)}(t)
	=
	\alpha_E
	\mathbf h_{\mathrm{BE}}^{(q)}
	(\mathbf h_{\mathrm{BE}}^{(q)})^{H}
	\mathbf w s(t)
	+
	\mathbf n^{(q)}(t),
\end{equation}
where $\alpha_E$ denotes the complex radar scattering coefficient,
which incorporates the target reflection property and propagation loss.
The radar cross section (RCS) fluctuation of Eve is modeled according to the Swerling-I model,
where
$\alpha_E\sim\mathcal{CN}(0,\sigma_E^2)$.
The scattering coefficient $\alpha_E$ is assumed to remain unchanged within
one coherent processing interval (CPI) and vary independently between
different CPIs~\cite{1057561,11230338,10564104}. Therefore, $\alpha_E$ is regarded as an unknown deterministic
parameter during the angular estimation process.
The RIS sequentially applies known phase configurations
$
\mathbf\Phi^{(q)}
=
{\rm diag}
(
e^{j\varphi_1^{(q)}},
\ldots,
e^{j\varphi_N^{(q)}}
),
q=1,\ldots,Q 
$.
After matched filtering, the collected sensing snapshots are denoted as
\begin{equation}
	\mathbf Y
	=
	[
	\mathbf y^{(1)},
	\mathbf y^{(2)},
	\ldots,
	\mathbf y^{(Q)}
	].
\end{equation}

\subsection{Angular Parameter Estimation}

The objective of angular parameter estimation is to obtain the direction
information of Eve from both the BS and the RIS observations. By exploiting
the spatial diversity provided by the reconfigurable RIS phase
configurations, the BS can acquire multiple sensing observations with
different spatial signatures.
For the $q$-th sensing interval, the equivalent BS-Eve channel is expressed
as
$\mathbf h_{\mathrm{BE}}^{(q)}=\mathbf h_{dE}+\mathbf h_{rE}^{(q)}$,
where $\mathbf h_{dE}$ denotes the direct BS-Eve channel and
$\mathbf h_{rE}^{(q)}$ represents the RIS-assisted BS-Eve channel under the
$q$-th RIS phase configuration, given by
$\mathbf h_{rE}^{(q)}=\mathbf G_{BR}^{H}(\mathbf\Phi^{(q)})^{H}\mathbf h_{rE}$.
The received sensing signal after matched filtering can be directly expressed as
\begin{equation}
	\begin{aligned}
		\mathbf y^{(q)}
		=
		\alpha_E
		\Big(
		\mathbf h_{dE}\mathbf h_{dE}^{H}
		+
		\mathbf h_{dE}
		(\mathbf h_{rE}^{(q)})^{H}
		+
		\mathbf h_{rE}^{(q)}
		\mathbf h_{dE}^{H}
		+
		\mathbf h_{rE}^{(q)}
		(\mathbf h_{rE}^{(q)})^{H}
		\Big)
		\mathbf w
		+
		\mathbf n^{(q)} .
	\end{aligned}
\end{equation}
The first term corresponds to the RIS-independent direct BS-Eve-BS
echo, while the remaining terms contain the RIS-dependent echo information.
By exploiting orthogonal RIS phase coding sequences, such as Discrete Fourier Transform (DFT)-based phase patterns, the RIS-independent and RIS-dependent components can be decoupled
from the received sensing snapshots~\cite{8937491}. Hence, the collected sensing data are
represented as
\begin{equation}
	\mathbf Y
	=
	\mathbf Y_d
	+
	\mathbf Y_r
	+
	\mathbf N ,
\end{equation}

\subsubsection{BS-Eve Direction Estimation}
After RIS code-based separation, the direct echo component can be expressed
as
\begin{equation}
	\mathbf Y_d
	=
	\alpha_E
	(\mathbf h_{dE}^{H}\mathbf w)
	\mathbf h_{dE}
	\mathbf 1_Q^T
	+
	\mathbf N_d .
\end{equation}
The direct BS-Eve channel follows the array response model
$\mathbf h_{dE}=\eta_{dE}\mathbf a_{\mathrm B}(\theta_{BE},\phi_{BE})$,
where $\eta_{dE}$ is the unknown complex channel coefficient and
$(\theta_{BE},\phi_{BE})$ denotes the azimuth and elevation angles of Eve
observed from the BS.
The spatial covariance matrix of the direct echo is estimated as
$\mathbf R_d=\frac{1}{Q}\mathbf Y_d\mathbf Y_d^{H}$.
By performing eigenvalue decomposition of $\mathbf R_d$, the noise
subspace matrix $\mathbf U_{n,d}$ is obtained. Based on the MUSIC algorithm~\cite{Krim1996,1143830}, the spatial spectrum is given by
\begin{equation}
	P_{\mathrm{BS}}(\theta,\phi)
	=
	\frac{1}
	{
		\mathbf a_{\mathrm B}^{H}(\theta,\phi)
		\mathbf U_{n,d}
		\mathbf U_{n,d}^{H}
		\mathbf a_{\mathrm B}(\theta,\phi)
	}.
\end{equation}
Accordingly, the BS-Eve direction estimate is obtained by searching for the
maximum of the MUSIC spectrum
\begin{equation}
	(\hat\theta_{BE},\hat\phi_{BE})
	=
	\arg\max_{\theta,\phi}
	P_{\mathrm{BS}}(\theta,\phi).
\end{equation}

\subsubsection{RIS-Eve Direction Estimation}

The RIS-dependent echo contains additional angular information observed
from the RIS. The RIS-Eve channel is modeled as
$\mathbf h_{rE}	=\eta_{rE}\mathbf a_{\mathrm R}(\theta_{RE},\phi_{RE})$,
where $\eta_{rE}$ denotes the unknown complex channel coefficient and
$(\theta_{RE},\phi_{RE})$ represents the direction of Eve with respect to
the RIS.
Let $\mathbf y_r^{(q)}$ denote the $q$-th RIS-dependent echo snapshot. It is
given by
\begin{equation}
	\begin{aligned}
		\mathbf y_r^{(q)}
		=
		\alpha_E
		\Big(
		\mathbf h_{dE}
		(\mathbf h_{rE}^{(q)})^H
		+
		\mathbf h_{rE}^{(q)}
		\mathbf h_{dE}^{H}
		+
		\mathbf h_{rE}^{(q)}
		(\mathbf h_{rE}^{(q)})^{H}
		\Big)
		\mathbf w
		+
		\mathbf n_r^{(q)} .
	\end{aligned}
\end{equation}
The first two terms correspond to the cross-path echoes between the direct
and RIS-assisted links, while the last term represents the RIS-mediated
round-trip echo.
According to the estimated BS-Eve direction, the direct channel is
constructed as
$\mathbf h_{dE}=\eta_{dE}\mathbf a_{\mathrm B}(\hat\theta_{BE},\hat\phi_{BE})$.
For a candidate RIS direction $(\theta,\phi)$, the normalized RIS response
is defined as
$\widetilde{\mathbf h}_{rE}^{(q)}(\theta,\phi)	=\mathbf G_{BR}^{H}(\mathbf\Phi^{(q)})^{H}\mathbf a_{\mathrm R}(\theta,\phi)$.
The unknown propagation-dependent coefficients are absorbed into three complex parameters
$
\beta_1=\alpha_E\eta_{dE}\eta_{rE}^{*},
\beta_2=\alpha_E\eta_{dE}^{*}\eta_{rE},
\beta_3=\alpha_E|\eta_{rE}|^2
$.
Therefore, the RIS-dependent echo model becomes
\begin{equation}
	\begin{aligned}
		\mathbf y_{\mathrm{mod}}^{(q)}
		=
		\beta_1
		\mathbf a_{\mathrm B}
		(\hat\theta_{BE},\hat\phi_{BE})
		\left(
		\widetilde{\mathbf h}_{rE}^{(q)}
		(\theta,\phi)
		\right)^H
		\mathbf w
		+
		\beta_2
		\left(
		\mathbf a_{\mathrm B}^{H}
		(\hat\theta_{BE},\hat\phi_{BE})
		\mathbf w
		\right)
		\widetilde{\mathbf h}_{rE}^{(q)}
		(\theta,\phi)
		+
		\beta_3
		\widetilde{\mathbf h}_{rE}^{(q)}
		(\theta,\phi)
		\left(
		\widetilde{\mathbf h}_{rE}^{(q)}
		(\theta,\phi)
		\right)^H
		\mathbf w .
	\end{aligned}
\end{equation}
Since $\mathbf G_{BR}$ and $\mathbf\Phi^{(q)}$ are known, the RIS-Eve
direction can be estimated by matching the measured RIS-dependent echoes
with the above physical model. Under the additive white Gaussian noise assumption, the MLE is formulated as
\begin{equation}
	\begin{aligned}
		(\hat\theta_{RE},
		\hat\phi_{RE},
		\hat\beta_1,
		\hat\beta_2,
		\hat\beta_3)
		=
		\arg\min_{\theta,\phi,\beta_1,\beta_2,\beta_3}
		\sum_{q=1}^{Q}
		\left\|
		\mathbf y_r^{(q)}
		-
		\mathbf y_{\mathrm{mod}}^{(q)}
		\right\|^2 .
	\end{aligned}
\end{equation}
For a fixed angular pair $(\theta,\phi)$, the above model is linear with
respect to the unknown complex coefficients
$\beta_1,\beta_2,\beta_3$. Therefore, these coefficients can be obtained by
least-squares estimation, which coincides with the MLE of the linear parameters under the Gaussian noise model, and the original optimization problem is reduced
to a two-dimensional angular search over $(\theta,\phi)$.
\subsection{Joint Angle-Based Position Estimation}

After obtaining the BS-Eve and RIS-Eve angular parameters, the Eve position
is estimated based on the geometric constraints provided by the BS and RIS.
Let $\mathbf p_B$, $\mathbf p_R$, and $\mathbf p_E$ denote the positions of
the BS, RIS, and Eve, respectively. The angular observation vector is defined as
$\boldsymbol{\vartheta}	=[\theta_{BE},\phi_{BE},\theta_{RE},\phi_{RE}]^T=
\mathbf f(\mathbf p_E)$,
where $\mathbf f(\cdot)$ represents the geometric relationship between the Eve position and the corresponding azimuth/elevation angles with respect to the BS and RIS. The estimated angular vector obtained from the MUSIC and maximum likelihood estimators is
$\hat{\boldsymbol{\vartheta}}=[\hat\theta_{BE},\hat\phi_{BE},\hat\theta_{RE},\hat\phi_{RE}]^T $.
Therefore, the Eve position is obtained by solving the following nonlinear least-squares problem
\begin{equation}
	\hat{\mathbf p}_E
	=
	\arg\min_{\mathbf p_E}
	\left\|
	\hat{\boldsymbol{\vartheta}}
	-
	\mathbf f(\mathbf p_E)
	\right\|_2^2 .
\end{equation}
By jointly exploiting the angular information from the BS and RIS, accurate
3D localization of Eve can be achieved.
\subsection{CRLB-Based Localization Accuracy Analysis}

The localization accuracy is characterized by the CRLB. The angular measurement model is expressed as
$\hat{\boldsymbol{\vartheta}}=\mathbf f(\mathbf p_E)+\Delta\boldsymbol{\vartheta}$,
where $\mathbf p_E=[x_E,y_E,z_E]^T$ denotes the Eve position,
$\mathbf f(\mathbf p_E)$ is the theoretical angle measurement function, and
$\Delta\boldsymbol{\vartheta}$ represents the angular estimation error.
Assuming a small-error Gaussian model,
$\Delta\boldsymbol{\vartheta}\sim\mathcal N(\mathbf0,\mathbf\Sigma_\vartheta)$,
the Fisher information matrix (FIM) with respect to Eve position is
$\mathbf J_p=\mathbf G_p^T\mathbf\Sigma_\vartheta^{-1}\mathbf G_p $,
where
$\mathbf G_p=\frac{\partial\mathbf f(\mathbf p_E)}{\partial\mathbf p_E}$
denotes the Jacobian matrix.
For the azimuth angle observed from node $i$ ($i\in\{B,R\}$),
\begin{equation}
	\nabla_{\mathbf p_E}\theta_i
	=
	\left[
	-\frac{y_E-y_i}{\rho_i^2},
	\frac{x_E-x_i}{\rho_i^2},
	0
	\right]^T ,
\end{equation}
where
$\rho_i^2=(x_E-x_i)^2+(y_E-y_i)^2 $.
For the elevation angle,
\begin{equation}
	\nabla_{\mathbf p_E}\phi_i
	=
	\left[
	-\frac{(z_E-z_i)(x_E-x_i)}
	{r_i^2\rho_i},
	-\frac{(z_E-z_i)(y_E-y_i)}
	{r_i^2\rho_i},
	\frac{\rho_i}{r_i^2}
	\right]^T ,
\end{equation}
where
$r_i^2=(x_E-x_i)^2+(y_E-y_i)^2+(z_E-z_i)^2 $.
According to the Cramér-Rao inequality,
${\rm cov}(\hat{\mathbf p}_E)\succeq\mathbf C_p $,
where the position CRLB is
$\mathbf C_p=\mathbf J_p^{-1}$.
Therefore, the CRLB-based localization accuracy is evaluated as the root mean square error (RMSE)
${\rm RMSE}_{\rm CRLB}=\sqrt{{\rm tr}(\mathbf C_p)} $.

\subsection{Delay-Aware IMM-EKF Tracking Model}
The radar sensing module yields 3D position estimates of the mobile Eve at discrete sensing instants after signal parameter estimation (e.g., via MLE or MUSIC algorithm). Due to sensing, processing, and transmission latencies, the obtained observation may correspond to a previous target state. The latency $d$ is assumed bounded and known after timestamping, which is practical in time-synchronized ISAC networks. Therefore, a buffer-based OOSM compensation model is introduced~\cite{BarShalom2005IMMOOSM}.
The target state is defined in a unified 10-dimensional space as
\begin{equation}
	\mathbf{x}_k = [x_k, y_k, z_k, v_{x,k}, v_{y,k}, v_{z,k}, a_{x,k}, a_{y,k}, a_{z,k}, \omega_k]^T \in \mathbb R^{10},
\end{equation}
where the first three elements form the 3D position vector $\mathbf p_k = [x_k, y_k, z_k]^T$, and the remaining components represent the 3D velocity vector, 3D acceleration vector, and horizontal yaw rate, respectively.
Considering a measurement delay of $d$ sampling intervals, the received observation is modeled as
\begin{equation}
	\mathbf z_k = \mathbf H \mathbf x_{k-d} + \mathbf v_k,
\end{equation}
where $\mathbf v_k \sim \mathcal N(\mathbf 0, \mathbf R)$ represents the measurement noise with measurement matrix $\mathbf H = [\mathbf I_{3\times3}, \mathbf 0_{3\times7}]$. To handle diverse maneuvering behaviors alongside delayed observations, an IMM framework incorporating constant velocity (CV), constant acceleration (CA), and coordinated turn (CT) modes is adopted. Model transitions follow a first-order Markov process with transition matrix elements $p_{ij} = \operatorname{Pr}(m_k=j | m_{k-1}=i)$.

\subsubsection{Model Initialization}
A short observation sequence $\mathcal D = \{(t_i, \mathbf p_i)\}_{i=1}^{N}$ containing preliminary position estimates is collected to initialize the motion models via least-squares fitting:

CV/CA Models: Linear/quadratic position polynomial fittings are performed independently across axes. The initial states in the unified space are constructed as \begin{align}
\mathbf x_0^{\mathrm{CV}} = [x_0, y_0, z_0, v_x, v_y, v_z, 0, 0, 0, 0]^T\\
\mathbf x_0^{\mathrm{CA}} = [x_0, y_0, z_0, v_x, v_y, v_z, a_x, a_y, a_z, 0]^T.
\end{align}

CT Model: Horizontal circular motion $(x_i - x_c)^2 + (y_i - y_c)^2 = R^2$ combined with vertical linear motion (assuming negligible vertical acceleration) is fitted to yield center $(x_c, y_c)$, radius $R$, and angular velocity $\omega$. Let $\theta$ be the initial phase angle; velocity and centripetal acceleration components are calculated as 
$v_x = -R\omega\sin\theta$, $v_y = R\omega\cos\theta$, $a_x = -R\omega^2\cos\theta$, and $a_y = -R\omega^2\sin\theta$ to form \begin{equation}
\mathbf x_0^{\mathrm{CT}} = [x_0, y_0, z_0, v_x, v_y, v_z, a_x, a_y, 0, \omega]^T.
\end{equation}

The Bayesian Information Criterion (BIC) values are calculated as
$
\mathrm{BIC}_{j}
=
N\ln
\left(
\frac{\mathrm{SSE}_{j}}{N}
\right)
+
r_j\ln N
$,
where $\mathrm{SSE}_{j}$ denotes the sum of squared errors between the
measured positions and the predicted positions under motion model $j$,
given by
$
\mathrm{SSE}_{j}
=
\sum_{k=1}^{N}
\left\|
\mathbf z_k-\hat{\mathbf z}_{k}^{j}
\right\|^2 
$.
Here, $r_j$ represents the number of parameters of model $j$, and $N$
is the number of observations used for model evaluation.
The BIC values are computed only to evaluate the relative fitting
performance among different motion models rather than directly
determining the IMM model probabilities. Therefore, the initial model
probabilities are uniformly set as
$\mu_0^j=1/3,\forall j\in\{\mathrm{CV},\mathrm{CA},\mathrm{CT}\}$.

\subsubsection{Delay-Aware IMM Filtering}
In the unified state space, the evolution for mode $j$ is given by
$
\mathbf{x}_k^j
=
\mathbf F_k^j\mathbf{x}_{k-1}^j
+
\mathbf u_k^j
$,
where
$\mathbf u_k^j\sim\mathcal{N}(\mathbf 0,\mathbf Q^j)$
denotes the process noise of mode $j$, and
$\mathbf Q^j$ is the corresponding process noise covariance matrix,
which characterizes the uncertainty caused by unmodeled dynamics and
random target maneuvers. The matrix $\mathbf F_k^j$ represents the
mode-dependent state transition matrix at step $k$.
For the nonlinear CT mode, the state evolution is linearized around the
predicted state and updated within the EKF
framework, where $\mathbf F_k^{\mathrm{CT}}$ denotes the corresponding
Jacobian matrix.

\subsubsection{Model Mixing \& Parallel Prediction}
At step $k$, the mixing probabilities are first calculated by
\begin{equation}
	\mu_{k-1}^{i|j}
	=
	\frac{p_{ij}\mu_{k-1}^{i}}
	{\sum_{l=1}^{3}p_{lj}\mu_{k-1}^{l}},
\end{equation}
and then used to obtain the mixed initial condition for each motion model:
\begin{equation}
	\bar{\mathbf{x}}_{k-1}^{j}
	=
	\sum_{i=1}^{3}
	\mu_{k-1}^{i|j}
	\mathbf{x}_{k-1}^{i},
\end{equation}
\begin{equation}
	\bar{\mathbf P}_{k-1}^{j}
	=
	\sum_{i=1}^{3}
	\mu_{k-1}^{i|j}
	\left[
	\mathbf P_{k-1}^{i}
	+
	(\mathbf x_{k-1}^{i}-\bar{\mathbf x}_{k-1}^{j})
	(\mathbf x_{k-1}^{i}-\bar{\mathbf x}_{k-1}^{j})^T
	\right].
\end{equation}
Each motion model then performs independent state prediction
\begin{equation}
	\hat{\mathbf{x}}_{k|k-1}^{j}
	=
	\mathbf F_k^{j}\bar{\mathbf{x}}_{k-1}^{j},
\end{equation}
\begin{equation}
	\mathbf P_{k|k-1}^{j}
	=
	\mathbf F_k^{j}
	\bar{\mathbf P}_{k-1}^{j}
	(\mathbf F_k^{j})^T
	+
	\mathbf Q^{j}.
\end{equation}
After receiving the measurement at step $k$, each model applies its
corresponding Kalman or extended Kalman measurement update to obtain the
posterior estimate
$\hat{\mathbf{x}}_{k|k}^{j}$ and covariance
$\mathbf P_{k|k}^{j}$.
The posterior states and covariances are then stored in the
mode-dependent finite buffer
\begin{equation}
	\mathcal B_k^j
	=
	\{
	\hat{\mathbf{x}}_{l|l}^{j},
	\mathbf P_{l|l}^{j}
	\}_{l=k-L}^{k},
	j\in\{\mathrm{CV},\mathrm{CA},\mathrm{CT}\},
\end{equation}
where $L$ denotes the maximum allowable measurement latency.

\subsubsection{OOSM Update \& Forward Propagation}
The OOSM compensation is performed independently for each motion hypothesis $j$ prior to model probability updates. Upon receiving a delayed measurement $\mathbf z_k$ corresponding to time $k-d$, the historical posterior state $\hat{\mathbf{x}}_{k-d|k-d}^{j}$ and covariance $\mathbf P_{k-d|k-d}^{j}$ are retrieved from $\mathcal B_k^j$. The innovation is $\tilde{\mathbf y}_{k-d}^{j} = \mathbf z_k - \mathbf H \hat{\mathbf{x}}_{k-d|k-d}^{j}$ with innovation covariance $\mathbf S_{k-d}^{j} = \mathbf H \mathbf P_{k-d|k-d}^{j} \mathbf H^T + \mathbf R$. 
Using Kalman gain $\mathbf K_{k-d}^{j} = \mathbf P_{k-d|k-d}^{j} \mathbf H^T (\mathbf S_{k-d}^{j})^{-1}$, the delayed state estimate at time $k-d$ is updated after incorporating the measurement received at time $k$ as
\begin{equation}
	\hat{\mathbf{x}}_{k-d|k}^{j} = \hat{\mathbf{x}}_{k-d|k-d}^{j} + \mathbf K_{k-d}^{j} \tilde{\mathbf y}_{k-d}^{j},
\end{equation}
\begin{equation}
	\begin{aligned}
		\mathbf P_{k-d|k}^{j} =  (\mathbf I - \mathbf K_{k-d}^{j}\mathbf H) \mathbf P_{k-d|k-d}^{j} (\mathbf I - \mathbf K_{k-d}^{j}\mathbf H)^T  + \mathbf K_{k-d}^{j} \mathbf R (\mathbf K_{k-d}^{j})^T .
	\end{aligned}
\end{equation}
After delay compensation, the corrected state is propagated forward to the current step $k$ via the $d$-step state transition operator $\mathcal F_j^{(d)}(\cdot)$ as follows
\begin{equation}
	\hat{\mathbf{x}}_{k|k}^{j} = \mathcal F_j^{(d)}\left(\hat{\mathbf{x}}_{k-d|k}^{j}\right),
\end{equation}
\begin{equation}
	\mathbf P_{k|k}^{j} = \mathbf \Phi_{k,k-d}^j \mathbf P_{k-d|k}^{j} (\mathbf \Phi_{k,k-d}^j)^T + \mathbf Q_{\mathrm{acc}}^{j},
\end{equation}
where $\mathbf \Phi_{k,l}^j$ denotes the equivalent state transition matrix from time $l$ to $k$. Specifically, $\mathbf \Phi_{k,k-d}^j$ equals $(\mathbf F_k^j)^d$ for linear CV/CA modes and the product of multi-step Jacobian matrices for the CT mode. The accumulated process noise covariance during the $d$-step forward prediction is expressed as
\begin{equation}
	\mathbf Q_{\mathrm{acc}}^{j} = \sum_{\ell=k-d+1}^{k} \mathbf\Phi_{k,\ell}^{j} \mathbf Q^{j} (\mathbf\Phi_{k,\ell}^{j})^T,
\end{equation}
where $\mathbf\Phi_{k,k}^{j} = \mathbf I$ represents the contribution of the process noise at the current filtering step.

\subsubsection{Likelihood Evaluation and Model Probability Update}
The delayed measurement likelihood is evaluated at the corresponding historical time instant, while the resulting model probability is updated at the current time index. The likelihood is given by
\begin{equation}
	\Lambda_k^{j} = \frac{\exp \left( -\frac12 (\tilde{\mathbf y}_{k-d}^{j})^T (\mathbf S_{k-d}^{j})^{-1} \tilde{\mathbf y}_{k-d}^{j} \right)}{\sqrt{(2\pi)^3 |\mathbf S_{k-d}^{j}|}}
\end{equation}
and is used to update the model probabilities as
\begin{equation}
	\mu_k^{j} = \frac{\Lambda_k^{j} \sum_{i=1}^{3} p_{ij}\mu_{k-1}^{i}}{\sum_{l=1}^{3} \Lambda_k^{l} \sum_{i=1}^{3} p_{il}\mu_{k-1}^{i}}.
\end{equation}

\subsubsection{Final State Fusion}
The overall IMM state estimate and error covariance are fused as
\begin{equation}
	\hat{\mathbf{x}}_{k|k} = \sum_{j=1}^{3} \mu_k^{j} \hat{\mathbf{x}}_{k|k}^{j},
\end{equation}
\begin{equation}
	\mathbf P_{k|k} = \sum_{j=1}^{3} \mu_k^{j} \Big[ \mathbf P_{k|k}^{j} + (\hat{\mathbf{x}}_{k|k}^{j} - \hat{\mathbf{x}}_{k|k}) (\hat{\mathbf{x}}_{k|k}^{j} - \hat{\mathbf{x}}_{k|k})^T \Big].
\end{equation}
The predicted Eve position, extracted via $\hat{\mathbf p}_{E,k} = \mathbf H \hat{\mathbf x}_{k|k} = [\hat{x}_k, \hat{y}_k, \hat{z}_k]^T$, is then used to update the time-varying Eve channel vector $\mathbf h_E(\hat{\mathbf p}_{E,k})$ for the subsequent secure transmission optimization.

\subsection{Computational Complexity Analysis}

The computational complexity of the proposed sensing and tracking framework mainly
comes from angular estimation and IMM tracking. For the BS-Eve angle estimation
using MUSIC, the covariance calculation, eigenvalue decomposition, and spectrum
search introduce complexities of $\mathcal{O}(M^2Q)$, $\mathcal{O}(M^3)$, and
$\mathcal{O}(G_{\mathrm{BS}}M^2)$, respectively. Thus, the overall complexity is
$\mathcal{O}(M^2Q+M^3+G_{\mathrm{BS}}M^2)$.
For the RIS-Eve angle estimation based on maximum likelihood fitting, the angular
search is performed over a two-dimensional grid. Considering the $Q$ sensing
snapshots and the BS-RIS channel processing, its complexity is
$\mathcal{O}(G_{\mathrm{RIS}}QMN)$.
The position estimation is a 3D nonlinear least-squares problem,
whose complexity is $\mathcal{O}(I_{\mathrm{pos}})$ with $I_{\mathrm{pos}}$
iterations. For the delay-aware IMM tracker, three motion models with state
dimension $n_x=10$ are processed in parallel, and the complexity is
$\mathcal{O}(N_m d n_x^3)$, where $N_m=3$ and $d$ denotes the measurement delay.

\section{Joint Transmit Beamforming and RIS Phase Shift Optimization Algorithm}

Based on the predicted location of the Eve provided by the
delay-aware IMM tracker, this section develops a joint transmit beamforming
and RIS phase shift optimization framework to maximize the secrecy rate
while satisfying the sensing accuracy requirement for reliable
Eve tracking.
The equivalent channels of Bob and Eve consist of the direct and
RIS-reflected links, which are expressed as
\begin{equation}
	\mathbf h_B
	=
	\mathbf h_{dB}
	+
	\mathbf G_{BR}^{H}
	\mathbf\Phi^{H}
	\mathbf h_{rB},
\end{equation}
\begin{equation}
	\mathbf h_E
	=
	\mathbf h_{dE}(\hat{\mathbf p}_{E,k})
	+
	\mathbf G_{BR}^{H}
	\mathbf\Phi^{H}
	\mathbf h_{rE}(\hat{\mathbf p}_{E,k}),
\end{equation}
The Eve-related channels are reconstructed according to the predicted
position $\hat{\mathbf p}_{E,k}$ obtained from the IMM tracker.
The received signals at Bob and Eve are
$
y_B=\mathbf h_B^H\mathbf w s+n_B
$,
$
y_E=\mathbf h_E^H\mathbf w s+n_E 
$.
The corresponding signal-to-interference-plus-noise ratios (SINRs) are
$
\gamma_B=
\frac{|\mathbf h_B^H\mathbf w|^2}{\sigma_B^2}
$,
$
\gamma_E=
\frac{|\mathbf h_E^H\mathbf w|^2}{\sigma_E^2}
$.
Therefore, the achievable secrecy rate is
\begin{equation}
	R_{\rm sec}
	=
	[
	\log_2(1+\gamma_B)
	-
	\log_2(1+\gamma_E)
	]^+ .
\end{equation}
The transmit waveform from the BS is jointly used for sensing and communication. During the sensing stage, the RIS phase configuration $\mathbf{\Phi}_{S}$ is designed to facilitate multi-view angular acquisition for Eve localization. After the sensing process is completed, the RIS phase shifts are re-optimized to meet the communication requirements. Therefore, the localization accuracy constraint is formulated as
$\operatorname{tr}(\mathbf C_p(\mathbf w))\leq\epsilon_{\rm loc}$.
The joint optimization problem is formulated as
\begin{subequations}
	\label{eq:opt_problem}
	\begin{align}
		\mathrm{P1:}\quad
		\max_{\mathbf w,\mathbf v} \quad & R_{\rm sec}(\mathbf w,\mathbf v)
		\nonumber \\
		\mathrm{s.t.} \quad & \|\mathbf w\|^2 \leq P_t,
		\label{eq:power_constraint} \\
		& |v_n| = 1, \quad n = 1,\ldots,N,
		\label{eq:unit_modulus} \\
		& \operatorname{tr}(\mathbf C_p(\mathbf w)) \leq \epsilon_{\rm loc}.
		\label{eq:localization_constraint}
	\end{align}
\end{subequations}

The above optimization problem is non-convex for the following three reasons. First, the secrecy rate is expressed as the difference of two logarithmic functions, resulting in a non-convex objective function. Second, the transmit beamformer $\mathbf{w}$ and the RIS phase shifts are coupled through the cascaded channel, giving rise to non-convex bilinear terms. Third, the unit-modulus constraints of the RIS elements, together with the CRLB constraint involving the inverse FIM, introduce additional sources of non-convexity.
To solve this problem, an AO framework is adopted,
where the transmit beamformer and RIS phase shift vector are optimized
alternatively.

\subsection{Transmit Beamforming Optimization with Fixed RIS Phase}

When the RIS phase shift vector $\mathbf v$ is fixed, the equivalent
channels of Bob and Eve become constants. Therefore, the beamforming
optimization problem is formulated as
\begin{subequations}
	\label{eq:opt_problem_p2}
	\begin{align}
		\mathrm{P2:}\quad
		\max_{\mathbf w} \quad & R_{\mathrm{sec}}(\mathbf w)
		\nonumber \\
		\mathrm{s.t.} \quad & \|\mathbf w\|^2 \leq P_t,
		\label{eq:power_constraint_p2} \\
		& \operatorname{tr}(\mathbf C_p(\mathbf w)) \leq \epsilon_{\mathrm{loc}}.
		\label{eq:localization_constraint_p2}
	\end{align}
\end{subequations}
Define $\mathbf H_B=\frac{\mathbf h_B\mathbf h_B^H}{\sigma_B^2}$, $\mathbf H_E=\frac{\mathbf h_E\mathbf h_E^H}{\sigma_E^2}$.
Since the logarithm function is monotonically increasing, maximizing the
secrecy rate is equivalent to maximizing the following fractional
objective:
\begin{equation}
	\max_{\mathbf w}
	\frac{
		1+\mathbf w^H\mathbf H_B\mathbf w
	}
	{
		1+\mathbf w^H\mathbf H_E\mathbf w
	}.
\end{equation}
The above fractional programming problem is solved by the Dinkelbach
method. For a given parameter $\eta^{(k)}$, the auxiliary problem is
expressed as
\begin{equation}
	\mathcal F(\eta^{(k)})
	=
	\max_{\mathbf w}
	\left\{
	1+\mathbf w^H\mathbf H_B\mathbf w
	-
	\eta^{(k)}
	(1+\mathbf w^H\mathbf H_E\mathbf w)
	\right\}.
\end{equation}
By defining
$
\mathbf A(\eta^{(k)})
=
\mathbf H_B-\eta^{(k)}\mathbf H_E 
$,
and ignoring the constant term $1-\eta^{(k)}$, the beamforming
subproblem becomes
\begin{subequations}
	\label{eq:opt_problem_p3}
	\begin{align}
		\mathrm{P3:}\quad
		\max_{\mathbf w} \quad & \mathbf w^H \mathbf A(\eta^{(k)}) \mathbf w
		\nonumber \\
		\mathrm{s.t.} \quad & \|\mathbf w\|^2 \leq P_t,
		\label{eq:power_constraint_p3} \\
		& \operatorname{tr}(\mathbf C_p(\mathbf w)) \leq \epsilon_{\mathrm{loc}}.
		\label{eq:localization_constraint_p3}
	\end{align}
\end{subequations}
The objective function is generally non-convex since
$\mathbf A(\eta^{(k)})$ is an indefinite matrix. Moreover, the CRLB
constraint contains the inverse of the FIM matrix and is
also non-convex. Therefore, successive convex approximation (SCA) and SDR techniques are adopted.
\subsubsection{SCA Approximation of CRLB Constraint}

Introducing the lifted matrix variable
$\mathbf W=\mathbf w\mathbf w^H $,
the CRLB metric can be equivalently represented as a function of
$\mathbf W$:
$\Psi(\mathbf W)=\operatorname{tr}(\mathbf C_p(\mathbf W))$.
At the $m$-th SCA iteration, the non-convex CRLB metric is locally
approximated around the previous point $\mathbf W^{(m)}$ as
\begin{equation}
	\Psi(\mathbf W)
	\approx
	\Psi_{\mathrm{lin}}^{(m)}(\mathbf W),
\end{equation}
where
$
	\Psi_{\mathrm{lin}}^{(m)}(\mathbf W)
	=
	\Psi(\mathbf W^{(m)})
	+
	\operatorname{tr}
	\left(
	\nabla_{\mathbf W}
	\Psi(\mathbf W^{(m)})^H
	(\mathbf W-\mathbf W^{(m)})
	\right)
$.
Here,
$\nabla_{\mathbf W}\Psi(\mathbf W^{(m)})$
denotes the Wirtinger gradient of the CRLB metric with respect to the
Hermitian matrix variable $\mathbf W$.
Accordingly, the original CRLB constraint is replaced by the following
linearized constraint:
$\Psi_{\mathrm{lin}}^{(m)}(\mathbf W)\leq\epsilon_{\mathrm{loc}}$.
\subsubsection{SDR-Based Beamforming Optimization}
Using the identity
$\mathbf w^H\mathbf A(\eta^{(k)})\mathbf w=\operatorname{tr}(\mathbf A(\eta^{(k)})\mathbf W)$
and temporarily removing the rank-one constraint
$\operatorname{rank}(\mathbf W)=1$, the SDR problem at the $m$-th SCA
iteration is formulated as
\begin{subequations}
	\label{eq:opt_problem_p4}
	\begin{align}
		\mathrm{P4:}\quad
		\max_{\mathbf W} \quad & \operatorname{tr}(\mathbf A(\eta^{(k)})\mathbf W)
		\nonumber \\
		\mathrm{s.t.} \quad & \operatorname{tr}(\mathbf W) \leq P_t,
		\label{eq:power_constraint_p4} \\
		& \Psi_{\mathrm{lin}}^{(m)}(\mathbf W) \leq \epsilon_{\mathrm{loc}},
		\label{eq:localization_constraint_p4} \\
		& \mathbf W \succeq 0.
		\label{eq:psd_constraint_p4}
	\end{align}
\end{subequations}
For a fixed Dinkelbach parameter $\eta^{(k)}$,
problem $\mathrm{(P4)}$ is solved iteratively using the SCA procedure.
The iteration continues until
$
\|\mathbf W^{(m+1)}
-
\mathbf W^{(m)}
\|
\leq
\epsilon_{\mathrm{SCA}} 
$.
After convergence of the SCA procedure, the obtained solution is denoted
as $\mathbf W^{(k)\star}$. Then, the Dinkelbach parameter is updated as
\begin{equation}
	\eta^{(k+1)}
	=
	\frac{
		1+
		\operatorname{tr}
		(
		\mathbf H_B
		\mathbf W^{(k)\star}
		)
	}
	{
		1+
		\operatorname{tr}
		(
		\mathbf H_E
		\mathbf W^{(k)\star}
		)
	}.\label{updata_eta}
\end{equation}
The Dinkelbach iteration terminates when
$
|\eta^{(k+1)}-\eta^{(k)}|
\leq
\epsilon_{\eta}
$.
\subsubsection{Beamforming Vector Recovery}
After convergence of the SDR-based optimization, the optimal matrix
$\mathbf W^\star$ is obtained. If
$\operatorname{rank}(\mathbf W^\star)=1 $,
the optimal beamforming vector can be directly recovered through
eigenvalue decomposition:
$\mathbf W^\star	=\lambda_1\mathbf u_1\mathbf u_1^H $,
where
$
\mathbf w^\star
=
\sqrt{\lambda_1}\mathbf u_1 
$.
Otherwise, Gaussian randomization is adopted. Specifically, $L$
candidate beamforming vectors are generated as
$
\tilde{\mathbf w}_l
=
\mathbf Q
\mathbf\Lambda^{1/2}
\mathbf z_l 
$,
where
$
\mathbf W^\star
=
\mathbf Q
\mathbf\Lambda
\mathbf Q^H 
$,
and
$
\mathbf z_l
\sim
\mathcal{CN}(\mathbf 0,\mathbf I)
$.
Each candidate vector is scaled to satisfy the transmit power constraint
\begin{equation}
	\tilde{\mathbf w}_l
	=
	\frac{\sqrt{P_t}}
	{\max(\|\tilde{\mathbf w}_l\|,\sqrt{P_t})}
	\tilde{\mathbf w}_l .
\end{equation}
The feasible candidate set is defined as
\begin{equation}
	\mathcal S_{\mathrm{feas}}
	=
	\left\{
	\tilde{\mathbf w}_l
	\left|
	\begin{array}{l}
		\|\tilde{\mathbf w}_l\|^2\leq P_t,
		\Psi(
		\tilde{\mathbf w}_l
		\tilde{\mathbf w}_l^H
		)
		\leq
		\epsilon_{\mathrm{loc}}
	\end{array}
	\right.
	\right\}.
\end{equation}
Finally, the optimal beamforming vector is selected as
\begin{equation}
	\mathbf w^\star
	=
	\arg\max_{\tilde{\mathbf w}_l\in\mathcal S_{\mathrm{feas}}}
	R_{\mathrm{sec}}
	(\tilde{\mathbf w}_l).
\end{equation}

\subsection{RIS Phase Shift Optimization via Riemannian Manifold Optimization}

With the optimized beamforming vector $\mathbf w$ fixed, the RIS phase shift optimization subproblem is formulated as
\begin{subequations}
	\label{eq:opt_problem_p5}
	\begin{align}
		\mathrm{P5:}\quad
		\min_{\mathbf v} \quad & J(\mathbf v) = -R_{\mathrm{sec}}(\mathbf v)
		\nonumber \\
		\mathrm{s.t.} \quad & |v_n| = 1, \quad n=1,\ldots,N .
		\label{eq:unit_modulus_p5}
	\end{align}
\end{subequations}
The unit-modulus constraint indicates that each RIS element belongs to the complex circle manifold. Therefore, the feasible set is defined as
$\mathcal M = \left\{\mathbf v\in\mathbb C^N : |v_n|=1,\,\forall n\right\}$.

\subsubsection{Equivalent Channel Representation}

For a fixed beamforming vector $\mathbf w$, the received signal components at Bob and Eve can be rewritten as $\mathbf h_B^H\mathbf w = c_B + \mathbf v^H\mathbf d_B$,
$\mathbf h_E^H\mathbf w = c_E + \mathbf v^H\mathbf d_E $,
where $c_B = \mathbf h_{dB}^{H}\mathbf w$, $c_E = \mathbf h_{dE}^{H}\mathbf w$, and the channel projection vectors are defined as
$\mathbf d_B = \operatorname{diag}(\mathbf h_{rB})\mathbf G_{BR}^{*}\mathbf w^{*},
\mathbf d_E = \operatorname{diag}(\mathbf h_{rE})\mathbf G_{BR}^{*}\mathbf w^{*}$.
The above definitions are mathematically consistent with the equivalent channels $\mathbf h_B = \mathbf h_{dB} + \mathbf G_{BR}^{H}\mathbf\Phi^{H}\mathbf h_{rB}$ and $\mathbf h_E = \mathbf h_{dE} + \mathbf G_{BR}^{H}\mathbf\Phi^{H}\mathbf h_{rE}$.
Accordingly, the received SINRs are expressed as
$\gamma_B(\mathbf v) = \frac{|c_B+\mathbf v^H\mathbf d_B|^2}{\sigma_B^2}, 
\gamma_E(\mathbf v) = \frac{|c_E+\mathbf v^H\mathbf d_E|^2}{\sigma_E^2}.$
Therefore, the objective function becomes
\begin{equation}
	J(\mathbf v) = -\log_2(1+\gamma_B(\mathbf v)) + \log_2(1+\gamma_E(\mathbf v)).
\end{equation}

\subsubsection{Riemannian Gradient Calculation}

Since $J(\mathbf v)$ is a real-valued function with respect to the complex variable $\mathbf v$, the Euclidean gradient is obtained using Wirtinger calculus. Defining $z_B = c_B + \mathbf v^H\mathbf d_B$ and $z_E = c_E + \mathbf v^H\mathbf d_E$, the Euclidean gradient with respect to $\mathbf v^*$ is given by
\begin{equation}
	\nabla_{\mathbf v^*}J(\mathbf v) = -\frac{1}{\ln2} \frac{\frac{z_B\mathbf d_B}{\sigma_B^2}}{1+\gamma_B} + \frac{1}{\ln2} \frac{\frac{z_E\mathbf d_E}{\sigma_E^2}}{1+\gamma_E}.
\end{equation}
For the complex circle manifold $\mathcal M$, the tangent space at point $\mathbf v$ is defined as
\begin{equation}
	T_{\mathbf v}\mathcal M = \left\{\boldsymbol\xi \in \mathbb{C}^N : \operatorname{Re}(\boldsymbol\xi\odot\mathbf v^*) = \mathbf 0 \right\}.
\end{equation}
The Euclidean gradient is projected onto the tangent space to obtain the Riemannian gradient:
\begin{equation}
	\operatorname{grad}J(\mathbf v) = \nabla_{\mathbf v^*}J(\mathbf v) - \operatorname{Re}\left\{ \nabla_{\mathbf v^*}J(\mathbf v) \odot \mathbf v^* \right\} \odot \mathbf v .
\end{equation}

\subsubsection{Riemannian Conjugate Gradient (RCG) Algorithm}

The RIS phase vector is updated along the Riemannian conjugate direction. At the $l$-th iteration, the search direction is constructed as
\begin{equation}
	\boldsymbol\eta^{(l)} = -\operatorname{grad}J(\mathbf v^{(l)}) + \beta^{(l)} \mathcal T_{\mathbf v^{(l-1)} \rightarrow \mathbf v^{(l)}}(\boldsymbol\eta^{(l-1)}),
\end{equation}
where $\mathcal T(\cdot)$ denotes the vector transport operation. For the complex circle manifold, the vector transport is efficiently implemented by projecting the previous search direction onto the tangent space of the current point $T_{\mathbf v^{(l)}}\mathcal M$.
The updated phase vector is obtained through the retraction operation:
\begin{equation}
	\mathbf v^{(l+1)} = \mathcal R_{\mathbf v^{(l)}}(\alpha^{(l)}\boldsymbol\eta^{(l)}).
\end{equation}
The step size $\alpha^{(l)}$ is determined by the Armijo backtracking line search on the manifold to guarantee a sufficient decrease of the objective function. For the complex circle manifold, the retraction operator is implemented element-wise as
\begin{equation}
	\mathcal R_{\mathbf v}(\boldsymbol\xi) = \exp\left( j\angle(\mathbf v+\boldsymbol\xi) \right).
\end{equation}
The RCG iteration terminates when $|J(\mathbf v^{(l+1)}) - J(\mathbf v^{(l)})| < \epsilon_{\mathrm{RIS}}$, where $\epsilon_{\mathrm{RIS}}$ denotes the convergence tolerance. Upon convergence, the optimized RIS phase shift vector is finally obtained as $\mathbf v^\star = \mathbf v^{(l+1)}$.

The overall joint optimization framework is summarized in Algorithm~\ref{alg:AO}.
\begin{algorithm}[t]
	\caption{Joint Transmit Beamforming and RIS Phase Shift Optimization Algorithm}
	\label{alg:AO}
	\begin{algorithmic}[1]
		
		\REQUIRE
		$\hat{\mathbf p}_{E,k},
		\mathbf h_{dB}, \mathbf h_{dE}, \mathbf h_{rB}, \mathbf h_{rE}, \mathbf G_{BR},
		P_t, \sigma_B^2, \sigma_E^2,
		\epsilon_{\mathrm{AO}}$, $\epsilon_{\eta}, \epsilon_{\mathrm{SCA}}, \epsilon_{\mathrm{RIS}}$.
		
		\ENSURE
		$\mathbf w^\star, \mathbf v^\star$.
		
		\STATE Initialize $\mathbf v^{(0)}$, set outer iteration index $i=0$.
		
		\REPEAT
		\STATE Fix $\mathbf v=\mathbf v^{(i)}$ and construct $\mathbf H_B, \mathbf H_E$.
		\STATE Initialize Dinkelbach parameter $\eta^{(0)}$, set $k=0$.
		
		\REPEAT
		\STATE Initialize SCA point $\mathbf W^{(0)}$, set $m=0$.
		\REPEAT
		\STATE Construct  $\Psi_{\mathrm{lin}}^{(m)}(\mathbf W)$.
		\STATE Solve SDP problem (P4) and obtain $\mathbf W^{(m+1)}$.
		\STATE Update $m = m + 1$.
		\UNTIL $\|\mathbf W^{(m)} - \mathbf W^{(m-1)}\| < \epsilon_{\mathrm{SCA}}$
		
		\STATE Set $\mathbf W^\star = \mathbf W^{(m)}$.
		\STATE Update
		$\eta^{(k+1)}$ via \eqref{updata_eta}.
		\STATE Update $k = k + 1$.
		\UNTIL $|\eta^{(k)} - \eta^{(k-1)}| < \epsilon_\eta$
		
		\STATE Recover $\mathbf w^{(i+1)}$ from $\mathbf W^\star$ using eigenvalue decomposition or Gaussian randomization.
		\STATE Fix $\mathbf w = \mathbf w^{(i+1)}$.
		\STATE Optimize $\mathbf v$ using the RCG algorithm on manifold $\mathcal M$.
		\STATE Obtain optimized RIS phase shift vector $\mathbf v^{(i+1)}$.
		\STATE Update $i = i + 1$.
		\UNTIL $|R_{\mathrm{sec}}^{(i)} - R_{\mathrm{sec}}^{(i-1)}| < \epsilon_{\mathrm{AO}}$
		
		\RETURN $\mathbf w^\star = \mathbf w^{(i)}, \mathbf v^\star = \mathbf v^{(i)}$.
		
	\end{algorithmic}
\end{algorithm}
\subsection{Computational Complexity Analysis}

The computational complexity of the proposed AO framework mainly stems from solving the active transmit beamforming and passive RIS phase shift optimization subproblems.
Let $I_{\mathrm{AO}}$, $I_D$, $I_{\mathrm{SCA}}$, and $I_{\mathrm{RG}}$ denote the iteration numbers of the outer AO loop, the Dinkelbach algorithm, the inner SCA procedure, and the RCG algorithm, respectively.
For a fixed RIS phase vector $\mathbf v$, solving the SDR problem $\mathrm{(P4)}$ via standard interior-point methods incurs a complexity of $\mathcal{O}(n_W^{3.5})$ per SCA iteration, where $n_W = M^2$ is the effective dimension of $\mathbf W$. Accounting for the nested Dinkelbach/SCA loops and the optional Gaussian randomization (requiring $\mathcal{O}(M^3 + LM^2)$), the beamforming subproblem complexity is
$\mathcal{O}_{\mathrm{BF}} = I_D I_{\mathrm{SCA}} \mathcal{O}(n_W^{3.5}) + \mathcal{O}(M^3 + LM^2)$.
For a fixed beamformer $\mathbf w$, calculating equivalent channel projection vectors and Euclidean gradients at each RCG iteration dominates with $\mathcal{O}(NM)$ operations. Thus, the RIS subproblem complexity is
$\mathcal{O}_{\mathrm{RIS}} = I_{\mathrm{RG}}\mathcal{O}(NM)$.
The total computational complexity is expressed as
$\mathcal{O}_{\mathrm{total}} = \mathcal{O}\left( I_{\mathrm{AO}} \left( I_D I_{\mathrm{SCA}} n_W^{3.5} + M^3 + LM^2 + I_{\mathrm{RG}}NM\right) \right)$.

\section{Numerical Results}
This section evaluates the performance of the proposed RIS-assisted secure ISAC scheme through numerical simulations. The impacts of the transmit power, the number of BS antennas, and the number of RIS elements on secrecy performance and target tracking accuracy are investigated. Unless otherwise specified, the simulation parameters are set as follows: the carrier frequency is $f_c=3.5$ GHz, the reference path loss coefficient is $C_0=-30$ dB, the Rician factor is $\mathcal K=10$, the system bandwidth is $B=20$ MHz, and the noise power for both communication and sensing is set to $-90$ dBm. The sensing duration occupies 0.2 of each coherent time slot, and the number of received snapshots is 1024. During the sensing stage, $L=4$ different RIS reflection configurations are employed to generate multiple spatial observations of the target. The BS employs a $4\times4$ UPA, the transmit power is set to 30 dBm, and the RIS consists of $4\times4$ reflecting elements.

The path loss exponents are set as $\alpha_{dB}=3.5$ for the BS-to-Bob link, $\alpha_{dE}=3.5$ for the BS-to-Eve link, $\alpha_{rB}=2.5$ for the RIS-to-Bob link, $\alpha_{rE}=2.5$ for the RIS-to-Eve link, and $\alpha_{BR}=2.2$ for the BS-RIS link. The spatial deployment is configured as follows: the BS is located at $\mathbf p_B=(0,0,5)$ m, the RIS is located at $\mathbf p_R=(20,20,10)$ m, the Bob is located at $\mathbf p_{Bob}=(25,0,5)$ m, and the initial position of the eavesdropper Eve is set as $\mathbf p_E=(28,5,5)$ m.

\begin{figure}[!h]
	\centering
	\includegraphics[width=4.5in]{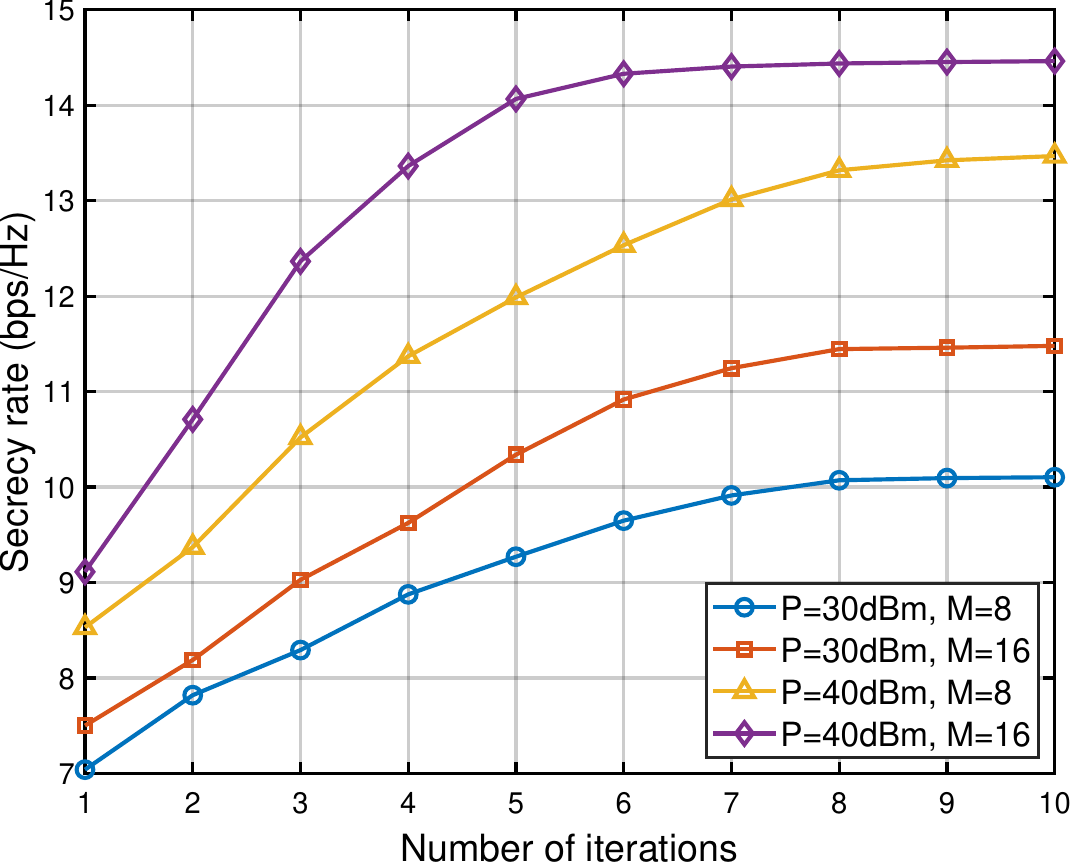}
	\caption{Convergence behavior of the proposed algorithm.}
	\label{Fig_2}
\end{figure}
As shown in Fig.\ref{Fig_2}, the proposed algorithm converges within approximately 10 iterations under different system configurations, which verifies the effectiveness and numerical stability of the adopted AO framework. As the transmit power increases, the secrecy rate improves significantly, with an increasingly pronounced enhancement. This is because higher transmit power provides sufficient power margins after satisfying the radar sensing constraint, allowing the beamforming design to more flexibly enhance the Bob signal and suppress the eavesdropper channel. Consequently, the achievable rate gap between the Bob and the Eve is enlarged, leading to an accelerated growth trend of the secrecy rate.

\begin{figure}[!h]
	\centering
	\includegraphics[width=4.5in]{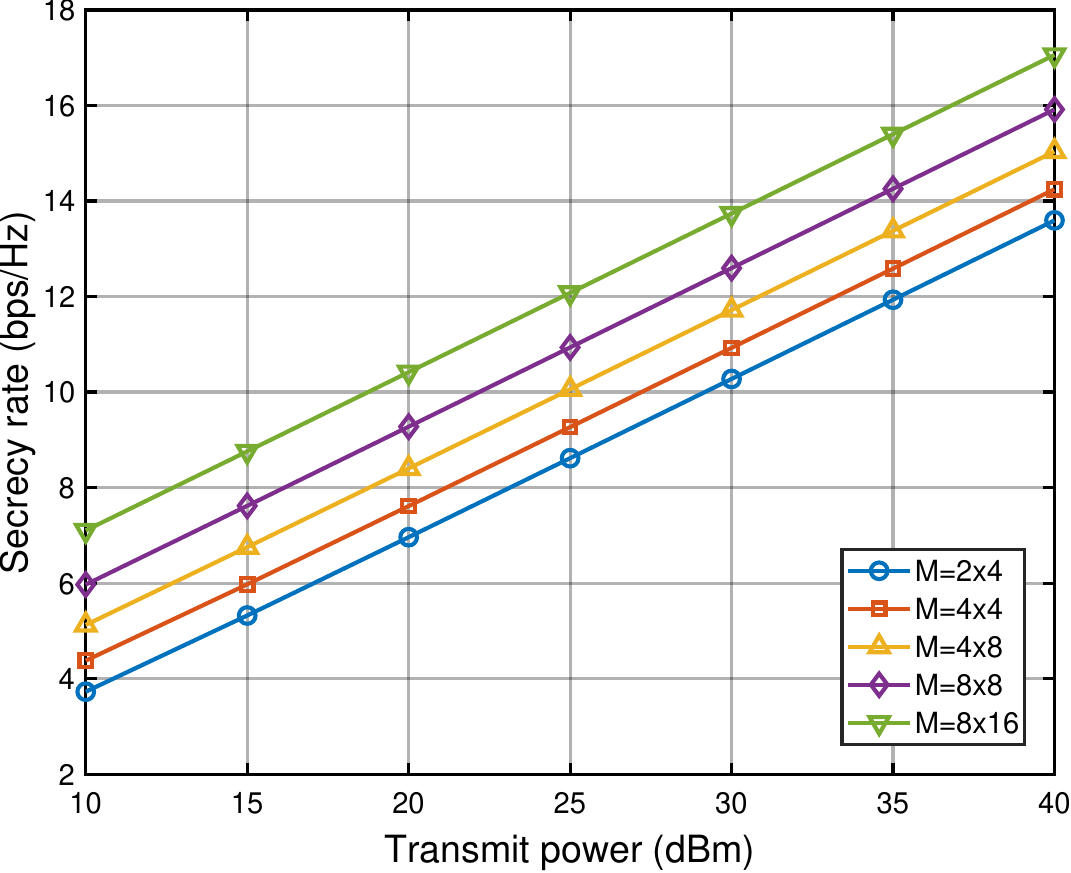}
	\caption{Effect of BS Array Size on the System Secrecy Rate.}
	\label{Fig_3}
\end{figure}
As shown in Fig.\ref{Fig_3}, the secrecy rate increases monotonically with the transmit power for all considered BS array configurations. Moreover, under the same transmit power, systems equipped with more antennas consistently achieve higher secrecy rates. This is because a larger antenna array provides higher spatial array gains and more degrees of freedom for beamforming design, enabling the BS to enhance the Bob link while suppressing the received signal power at the Eve under the sensing constraints. In addition, a larger array aperture improves the angular resolution during the sensing stage, thereby enhancing the localization accuracy of the Eve and providing more accurate state information for subsequent secure transmission optimization.

\begin{figure}[!h]
	\centering
	\includegraphics[width=4.5in]{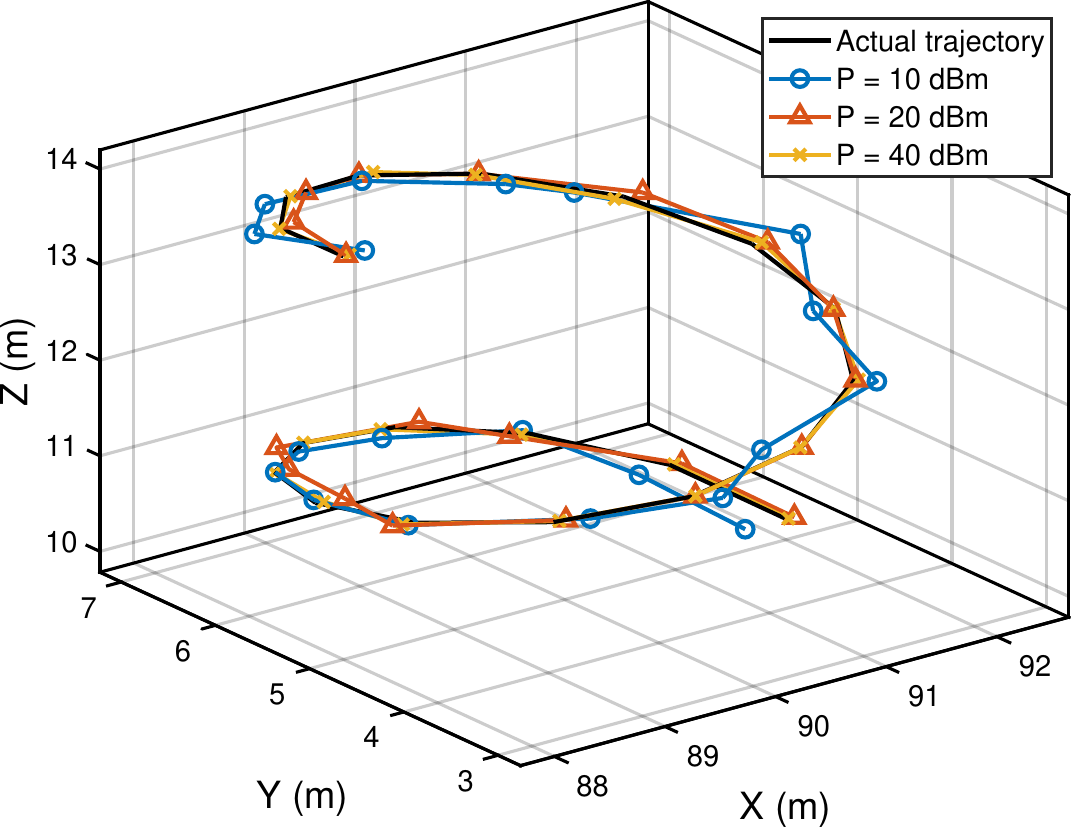}
	\caption{Comparison between true and estimated trajectories.}
	\label{trajectories}
\end{figure}
\begin{figure}[!h]
	\centering
	\includegraphics[width=4.5in]{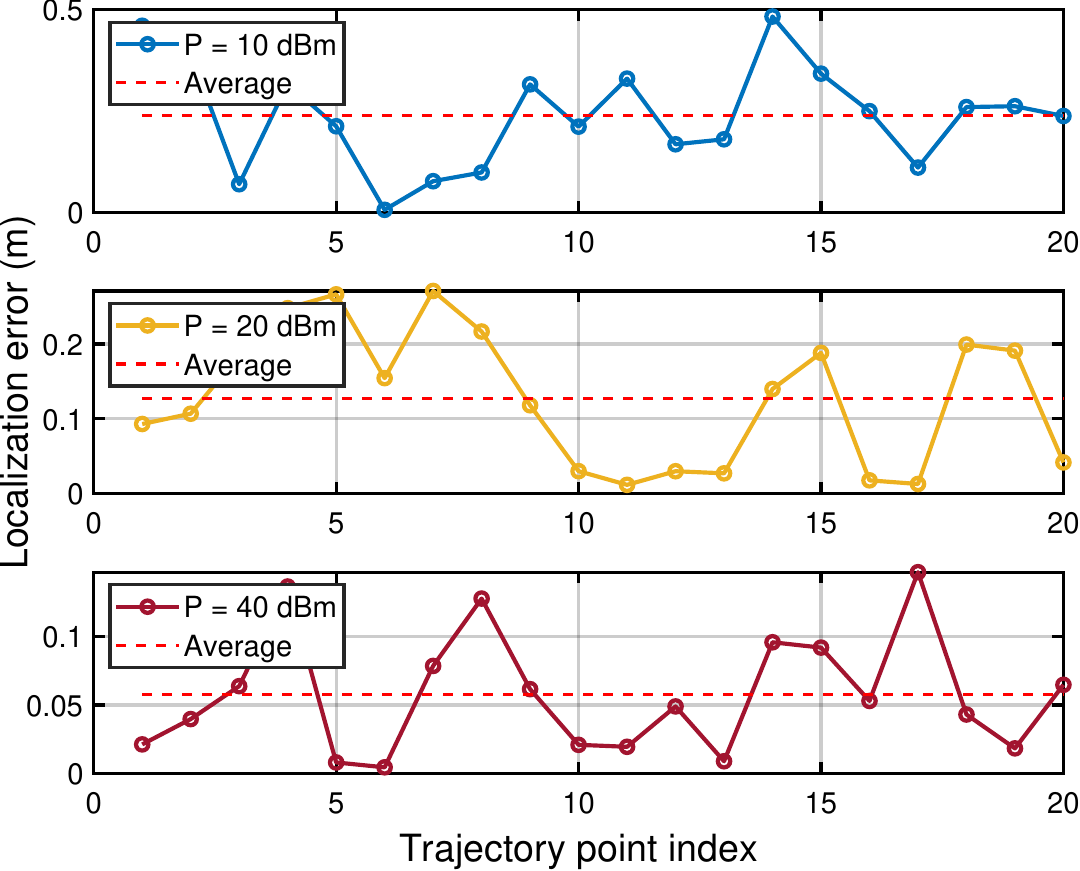}
	\caption{RMSE of trajectory estimation versus transmit power.}
	\label{fig_rmse_vs_power}
\end{figure}
The 3D trajectory estimation results based on radar echoes are presented in Fig.\ref{trajectories}. It can be observed that the estimated trajectory gradually approaches the ground-truth trajectory as the transmit power increases, indicating that the proposed sensing scheme can effectively acquire the target position information. Fig.\ref{fig_rmse_vs_power} further illustrates the relationship between the sensing error and transmit power, where the estimation error decreases monotonically with increasing transmit power, which is consistent with the theoretical expectation.

\begin{figure}[!h]
	\centering
	\includegraphics[width=4.5in]{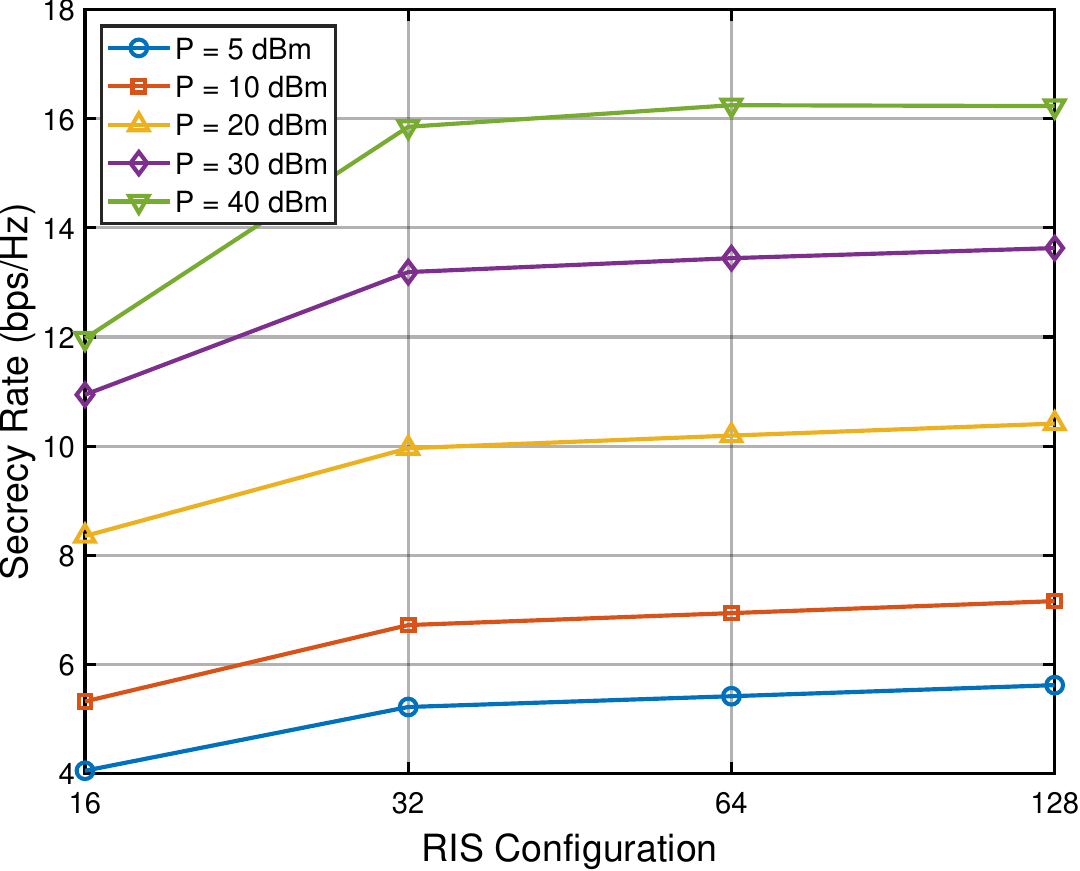}
	\caption{Secrecy rate versus the number of RIS elements.}
	\label{Fig_5}
\end{figure}
Fig.~\ref{Fig_5} illustrates the impact of the number of RIS elements on the secrecy rate. The results demonstrate that the secrecy performance improves as the RIS size increases. This is because a larger number of RIS reflecting elements can provide higher passive beamforming gain and finer spatial degrees of freedom, enabling the reflected signals to be more effectively focused toward the legitimate user Bob while suppressing the communication rate of the eavesdropper Eve, thereby effectively enhancing the secrecy transmission capability of the system.

\begin{figure}[!h]
	\centering
	\includegraphics[width=4.5in]{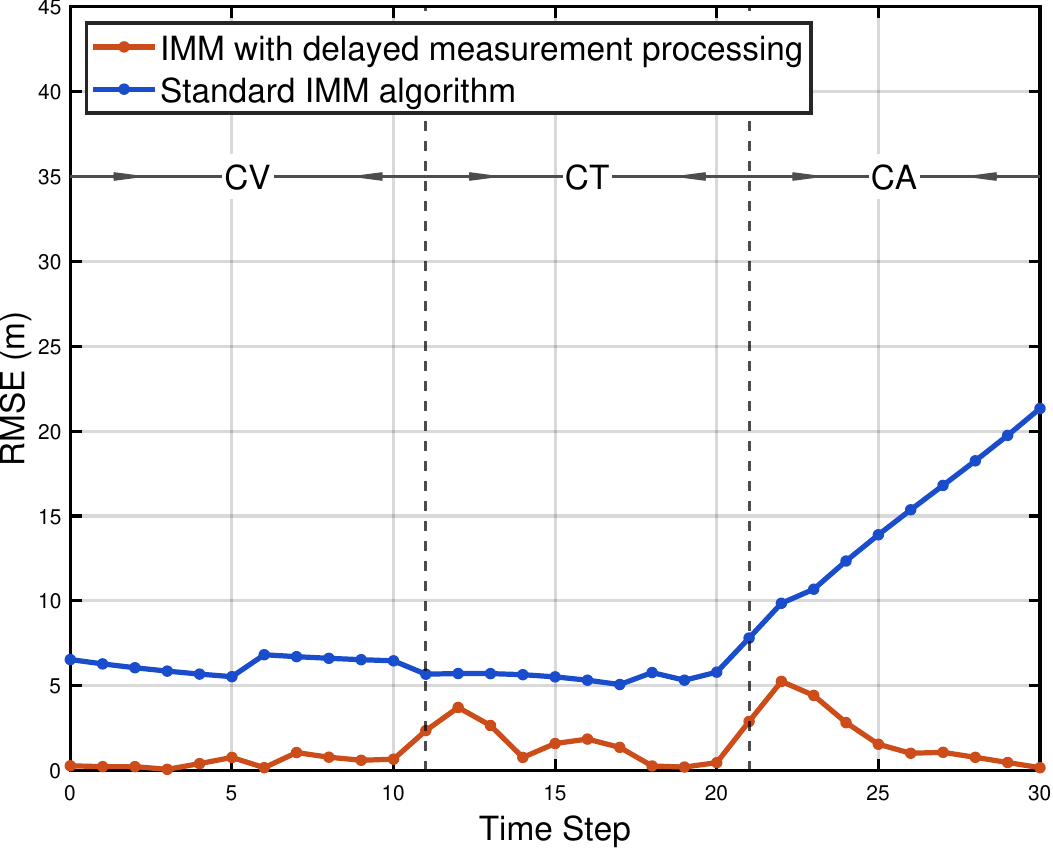}
	\caption{Performance comparison between delay-aware IMM and conventional IMM.}
	\label{Fig_8}
\end{figure}
Fig.~\ref{Fig_8} compares the RMSE performance of the proposed delay-aware IMM algorithm and the conventional IMM algorithm for maneuvering target tracking. The results demonstrate that the proposed method consistently achieves lower tracking errors throughout the entire observation period, validating the effectiveness of the delay compensation mechanism. It is worth noting that a slight performance degradation occurs during the initial stage of motion model switching due to the transient effects introduced by delayed updates and forward propagation. Nevertheless, the proposed method still provides significant improvements in overall estimation accuracy and tracking stability.

\section{Conclusion}
This paper investigated sensing driven secure transmission for RIS assisted ISAC systems with a mobile Eve. A multi view sensing and localization method was developed by exploiting multiple RIS phase configurations, and the corresponding 3D localization accuracy was characterized using the CRLB. To account for sensing delays and different Eve motion patterns, a delay aware IMM-EKF with OOSM compensation was employed to track and predict the Eve state. The predicted state was then used to update the time varying eavesdropping channel, based on which the BS beamforming and RIS phase shifts were jointly optimized under sensing and communication constraints. An AO algorithm was developed to solve the resulting non convex problem. Numerical results demonstrated that the proposed approach achieved accurate Eve tracking under different motion patterns and sensing delays and provided improved secrecy performance compared with schemes relying on static or outdated Eve information. Future work will consider multiple mobile Eves and learning based trajectory prediction.

\balance
\bibliographystyle{IEEEtran}  
\bibliography{ref}     

\end{document}